\documentclass[aps,prl,twocolumn,showpacs,showkeys,floatfix,footinbib,longbibliography,superscriptaddress]{revtex4-1}
\usepackage{amsmath}
\usepackage{amssymb}
\usepackage{graphicx}
\usepackage{epstopdf}
\usepackage{bm}
\usepackage{color}
\usepackage{relsize}
\usepackage{braket}
\usepackage[caption=false]{subfig}
\usepackage{lipsum}
\usepackage{hyperref}
\usepackage{fixltx2e}
\usepackage{amsfonts}
\usepackage{bbm}
\usepackage[all]{hypcap}

\usepackage{amssymb}
\usepackage{mathrsfs}
\usepackage{amsthm}

\begin{document}
\title{Enhanced spin-triplet pairing in magnetic junctions with s-wave superconductors}
\author{Thomas Vezin}
\affiliation{Department of Physics, University at Buffalo, State University of New York, Buffalo, NY 14260, USA}
\affiliation{Laboratoire des Solides Irradies, Ecole Polytechnique, Universite Paris-Saclay, F-91767 Palaiseau Cedex, France}
\author{Chenghao Shen}
\affiliation{Department of Physics, University at Buffalo, State University of New York, Buffalo, NY 14260, USA}
\author{Jong E. Han}
\affiliation{Department of Physics, University at Buffalo, State University of New York, Buffalo, NY 14260, USA}
\author{Igor \v{Z}uti\'c}
\email{zigor@buffalo.edu}
\affiliation{Department of Physics, University at Buffalo, State University of New York, Buffalo, NY 14260, USA}
\affiliation{Laboratoire des Solides Irradies, Ecole Polytechnique, Universite Paris-Saclay, F-91767 Palaiseau Cedex, France}

\begin{abstract}
A common path to superconducting spintronics, Majorana fermions, and topologically-protected quantum computing relies on spin-triplet superconductivity.  While naturally occurring spin-triplet pairing is elusive and even common spin-triplet candidates, such as Sr$_2$RuO$_4$, support alternative explanations, proximity effects in heterostructures can overcome these limitations. It is expected that robust spin-triplet superconductivity in magnetic junctions should rely on highly spin-polarized magnets or complex magnetic multilayers. Instead, we predict that the interplay of interfacial spin-orbit coupling and the barrier strength in simple magnetic junctions, with only a small spin polarization and s-wave superconductors, can lead to nearly complete spin-triplet superconducting proximity effects. This peculiar behavior arises from an effective perfect transparency: interfacial spin-orbit coupling counteracts the native potential barrier for states of a given spin and wave vector. We show that the enhanced spin-triplet regime is characterized by a huge increase in conductance magnetoanisotropy, orders of magnitude larger than in the normal state.
\end{abstract}
\maketitle

Realizing equal-spin triplet superconductivity provides an important platform for implementing superconducting spintronics and topologically-protected Majorana bound states (MBS)~\cite{Eschrig2011:PT,Linder2015:NP,Martinez2016:PRL,Aasen2016:PRX,Kitaev2001:PU,Mourik2012:S,Rokhinson2012:NP}. While naturally occurring triplet pairing remains elusive, transforming materials through proximity effects~\cite{Zutic2019:MT} offers a promising path to tailor the desired superconducting pairing~\cite{Kontos2001:PRL,Ryazanov2001:PRL,Fu2008:PRL,Buzdin2005:RMP,Bergeret2005:RMP}. 

For superconducting spintronics equal-spin triplet supports pure spin currents and 
the coexistence of superconductivity and ferromagnetism through long-range superconducting proximity effects in ferromagnet/superconductor (F/S) 
junctions~\cite{Buzdin2005:RMP,Bergeret2005:RMP,Eschrig2015:RPP}.  
Such junctions typically rely on multiple ferromagnetic and 
superconducting regions~\cite{Buzdin2005:RMP,Bergeret2005:RMP,Gingrich2016:NM,Cottet2011:PRL,Khaire2010:PRL}, 
complex ferromagnets with spiral magnetization~\cite{Robinson2010:S}, 
or complete spin polarization in half-metallic ferromagnets~\cite{Singh2015:PRX,Alidoust2018:PRB,Keizer2006:N},

With alternative paths towards spin-triplet pairing, where interfacial spin-orbit coupling (SOC) could relax the requirement of a complex magnetic structure, 
it is expected that both a strong spin polarization and strong SOC are needed~\cite{Jeon2018:NM,Banerjee2018:PRB,Satchell2018:PRB,Johnsen2019:P}.
However, we reveal that for nearly complete spin-triplet proximity-induced superconductivity even weakly spin-polarized ferromagnet and smaller SOC could be desirable. Our findings could complement the paths towards MBS where proximity-induced spin-triplet pairing is sought through strong SOC 
and half-metallic ferromagnets~\cite{Fu2008:PRL,Duckheim2011:PRB,Lutchyn2010:PRL,Oreg2010:PRL}.

A microscopic understanding of a superconducting proximity effect is obtained from the process of Andreev reflection (AR)
at interfaces with superconductors 
where  an electron is reflected backwards and converted into a hole with opposite charge and spin. This implies the doubling of the normal state 
conductance~\cite{Blonder1982:PRB} since two electrons are transferred across the interface into the S region where they form a spin-singlet Cooper pair.
In contrast to this conventional AR,
a spin-active interface with interfacial spin-flip scattering also yields AR
with an equal spin of electrons and holes~\cite{Zutic1999:PRBa}, responsible for a spin-triplet Cooper pair.

\begin{figure}[t]
\centering
\includegraphics*[trim=0.8cm 0.6cm 0.9cm 0.5cm,clip,width=8.5cm]{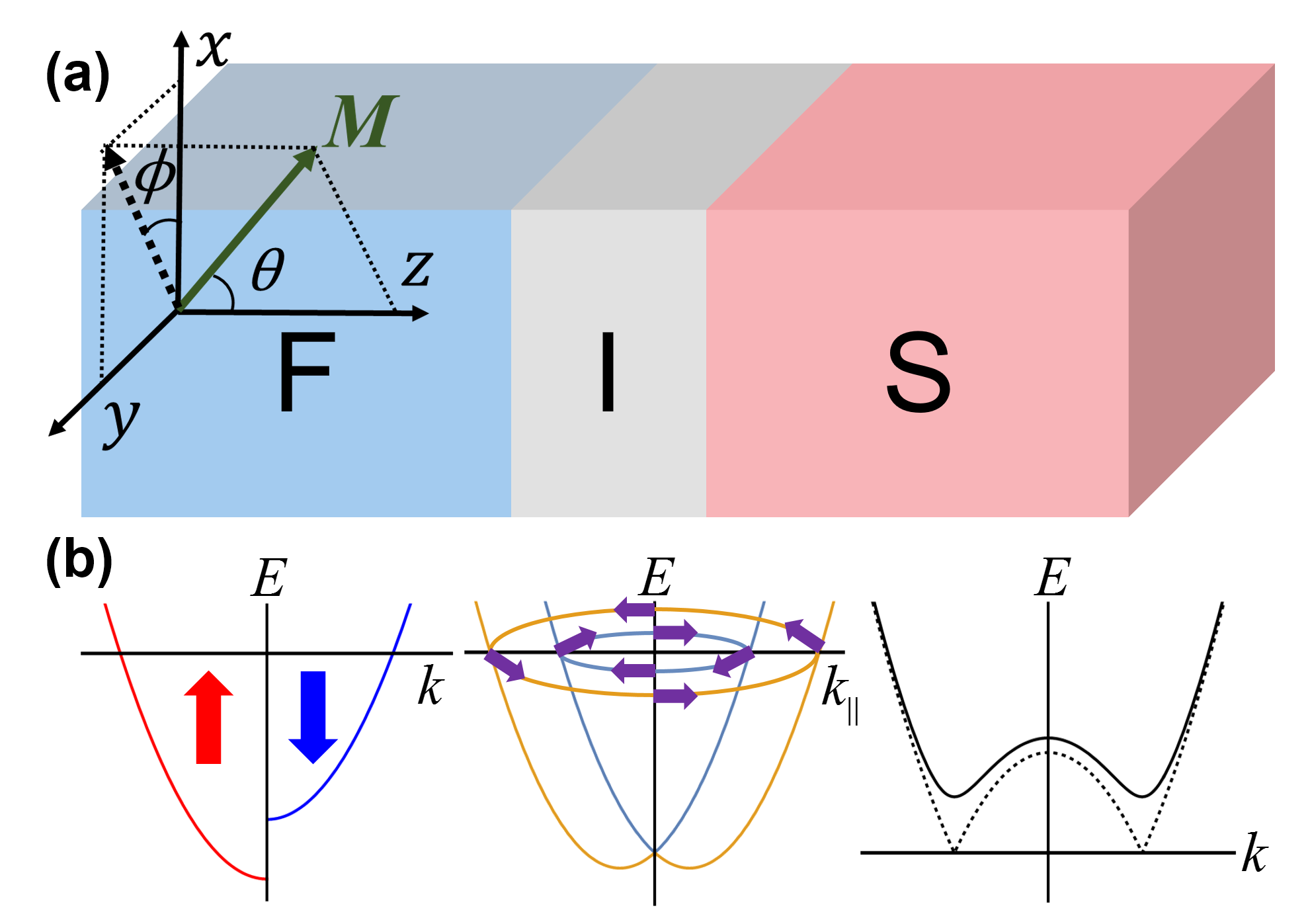}
\caption{
Ferromagnet/superconductor (F/S) junction, F and S separated by a flat interface (I) with potential and Rashba spin-orbit scattering (SOC).
$\bm{M}$ is the magnetization and the current flows normal to I. (b) Schematic band structure in each region. Spin are denoted by arrows:
In the F region red (blue) for parallel (antiparallel) to $\bm{M}$; 
with interfacial SOC, spins are 
parallel to the interface and $\perp$ to the in-plane component of the momentum, $k_\parallel$.
Excitation picture in the S region, the dashed line shows the normal state dispersion.
}	
\label{fig:TF1}
\end{figure}

We consider F/S junction, depicted in Fig.~\ref{fig:TF1},  
having a flat interface (I) at $z=0$ with potential and Rashba SOC scattering~\cite{Zutic2004:RMP}. 
We generalize the Blonder-Tinkham-Klapwijk formalism~\cite{Blonder1982:PRB,Kashiwaya2000:RPP,Granstrom2017:P}
to solve Bogoliubov-de Gennes equation for quasiparticle states $\Psi \left( \bm{r} \right)$  with energy E~\cite{Hogl2015:PRL},
\begin{equation} 
\left( {\begin{array}{*{20}{c}}
	{{{\hat H}_e}}&{\Delta \Theta (z){I_{2 \times 2}}} \\ 
	{{\Delta ^*}\Theta (z){I_{2 \times 2}}}&{{{\hat H}_h}} 
	\end{array}} \right)\Psi \left(\bm{r} \right) = E \Psi\left(\bm{r}\right), 
	\label{eq:H}
\end{equation}
where the single-particle Hamiltonian for electrons 
is $\hat{H}_e=-(\hbar^2/2) \boldsymbol{\nabla}\left[1/m(z)\right]\boldsymbol{\nabla}-\mu(z)- (\Delta_{xc}/2) 
\Theta(-z) \mathbf{m}\cdot\boldsymbol{\hat{\sigma}}+ [V_0 d + \alpha (k_y {\hat \sigma }_x - k_x{\hat \sigma}_y)]\delta(z)$ 
and  for holes $\hat{H}_h=-\hat{\sigma}_y\hat{H}_e^*\hat{\sigma}_y$. 
They contain the effective mass $m(z)$, 
the chemical potential $\mu(z)$, and the exchange spin splitting $\Delta_{xc}$. Magnetization, $\bm{M}$, has
orientation $\bm{m} = \left( {\sin \theta \cos  \phi , \sin \theta \sin \phi , \cos \theta } \right)$,
$\boldsymbol{\hat{\sigma}}$ are Pauli matrices, and $\mathbf{k}$ is wave vector. The interfacial scattering is modeled by delta-like potential barrier with effective height $V_0$ and width $d$ and the Rashba SOC with 
strength $\alpha$, due to structure inversion asymmetry~\cite{Zutic2004:RMP}.
The s-wave superconductor is described by the constant pair potential $\Delta$.

Since the in-plane wave vector $\bm{k}_\parallel$ is conserved, the scattering states for incident spin $\sigma$ electron are given by $\Psi_\sigma \left(\bm{r} \right) = e^{i\bm{k}_\parallel \cdot \bm{r}_\parallel}   \psi_\sigma \left(z \right)$ in a four-component basis~\cite{Zutic1999:PRBa} 
where the ``bar" symbol denotes the spin-flip contribution
\begin{widetext}
\begin{equation} 
{\psi _\sigma }\left( z \right) = \left\{ {\begin{array}{*{20}{l}}
	{\chi _\sigma ^e{e^{ik_\sigma ^ez}} + {a_\sigma }\chi _{-\sigma} ^h{e^{ik_{ - \sigma }^hz}} + {b_\sigma }\chi _\sigma ^e{e^{ - ik_\sigma ^ez}} + {\bar{a}_{\sigma }}\chi _\sigma ^h{e^{ik_\sigma ^hz}} + {\bar{b}_{\sigma }}\chi _{ - \sigma }^e{e^{ - ik_{ - \sigma }^ez}}}&{{\text{for}}\;z < 0,} \\ 
	{{c_\sigma }\left( {\begin{array}{*{20}{c}}
			u \\ 0 \\ v \\ 0 
			\end{array}} \right){e^{i{q^e}z}} + {d_\sigma }\left( {\begin{array}{*{20}{c}}
			v \\ 0 \\ u \\ 	0 
			\end{array}} \right){e^{ - i{q^h}z}} + {\bar{c}_{\sigma }}\left( {\begin{array}{*{20}{c}}
			0 \\ u \\ 0 \\ 	v 
			\end{array}} \right){e^{i{q^e}z}} + {\bar{d}_{\sigma }}\left( {\begin{array}{*{20}{c}}
			0 \\ v \\ 0 \\ u 
			\end{array}} \right){e^{ - i{q^h}z}}}&{{\text{for}}\;z > 0}. 
	\end{array}} \right.
	\label{eq:WF}
\end{equation}
\end{widetext}

In the F region, the eigenspinors for electrons and holes are 
$\chi _\sigma^e = {\left( {{\chi _\sigma },0} \right)^T}$ and $\chi _\sigma ^h = {\left( {0,{\chi _{ - \sigma }}} \right)^T}$ with
\begin{equation}\label{spinor}
{\chi _\sigma} = (1/\sqrt{2}){\left( {\sigma \sqrt {1 + \sigma \cos \theta } {e^{ - i\phi }},\sqrt{1 - \sigma \cos \theta }} \right)}^T,
\end{equation}
where $\sigma=1 (-1)$ refer to spin parallel (antiparallel) to $\bm{M}$ and the $z$-components of the 
wave vector are
$k^{e\,(h)}_\sigma = \sqrt{k_F^2+(2m_F/\hbar^2)\left[(-)E + \sigma \Delta_{xc}/2\right]-k_\parallel^2}$, 
with a spin-averaged
Fermi wave vector, $k_F$~\cite{Zutic2000:PRB}. In the S region, coherence factors, $u$, $v$, satisfy
 ${u^2} = 1 - {v^2} = \left( {1 + \sqrt {{E^2} - {\Delta ^2}} /E} \right)/2$, while the $z$-components of the 
 wave vector are
 $q^{e\,(h)} = \sqrt{q_F^2+(-)(2m_S/\hbar^2)\sqrt{E^2-\Delta^2}-k_\parallel^2}$, with $q_F$ the Fermi wave vector. 
 Similar to Snell's law~\cite{Zutic2000:PRB}, for a large $k_\parallel$ these $z$-components can become imaginary
 representing evanescent states which carry no net current. 

\begin{figure}[b]
\centering
\includegraphics*[trim=0.3cm 0.95cm 0.5cm 1.73cm,clip,width=8.5cm]{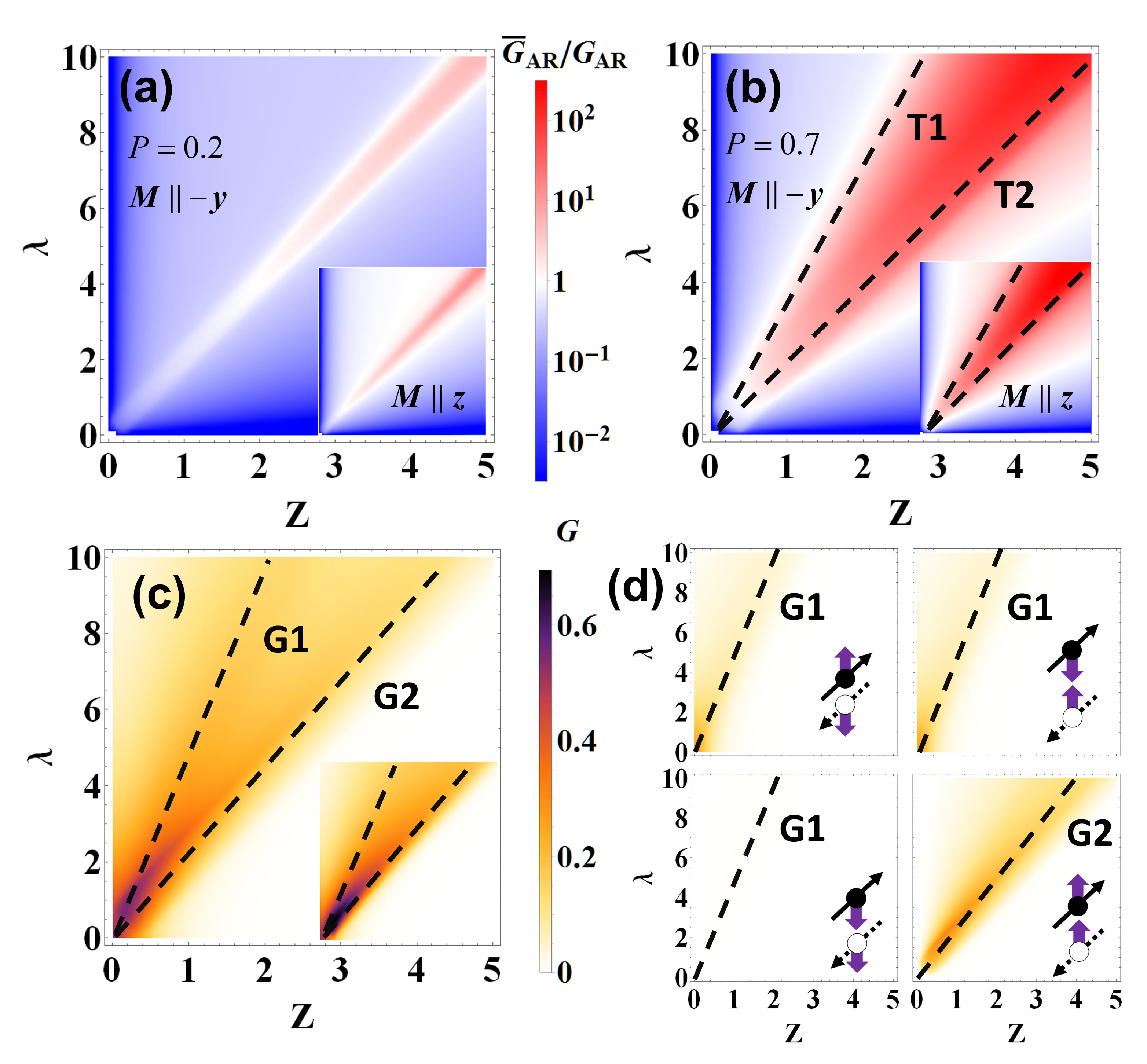} 
\caption{
The conductance ratio between the spin-flip and conventional Andreev reflection as a function of barrier potential $Z$ and Rashba SOC strength 
$\lambda$ for spin polarization (a) $P = 0.2$, (b) $P = 0.7$ (b) with in-plane $\bm{M}$. The insets: out-of-plane $\bm{M}$. 
(c) The total conductance as a function $Z$ and $\lambda$ for in-plane and out-of-plane (inset) $\bm{M}$ with $P = 0.7$
and (d)  its contributions from different processes, solid (dashed) arrows: incoming electrons (reflected holes), 
violet arrows: spin parallel (up) and antiparallel (down) to $\bm{M}$.
}
\label{fig:TF2}
\end{figure}

From the charge current conservation, we can express zero-temperature conductance at applied bias, $V$,
\begin{equation}
G(V) =  \sum\limits_\sigma  \int \frac{d k_\parallel}{2 \pi k_F^2} \left[ 1 + R_\sigma^h(- eV) - R_\sigma^e(eV) \right],
\label{eq:G}
\end{equation}
normalized by the Sharvin conductance $G_\mathrm{Sh} = e^2 k_F^2 A/(2 \pi h)$ \cite{Zutic2004:RMP}, where $A$ is the interfacial area.
Only the probability amplitudes from the F region are needed,  
for  Andreev  $R_\sigma^h = \operatorname{Re} [(k_{-\sigma}^h/k_\sigma^e){\left| a_\sigma \right|^2} + (k_\sigma^h/ k_\sigma^e){\left| {\bar{a}_\sigma} \right|^2}]$ and specular reflection $R_\sigma^e = \operatorname{Re} [{\left| b_\sigma \right|^2 + (k_{-\sigma}^e/k_\sigma^e) \left| \bar{b}_\sigma \right|^2}]$.

We focus on the zero-bias conductance, $G(0)$, where there is no quasiparticle transmission and, from the 
probability conservation~\cite{Zutic1999:PRBa,Hogl2015:PRL}, can be expressed using AR such that in Eq.~(\ref{eq:G}) the integration kernel 
is $2[R_\sigma^h(0)]$. The total conductance can be decomposed into four processes: conventional and spin-flip AR for spin-up (spin-down) 
$\uparrow$ ($\downarrow$) incident electrons corresponding, respectively, to the spin-singlet and spin-triplet superconducting correlations at the interface. 
It is convenient to introduce spin polarization $P=\Delta_{xc}/2\mu_F$, and dimensionless parameters for barrier strength 
$Z =V_0 d \sqrt{m_F m_S}/(\hbar^2 \sqrt{k_F q_F})$ and Rashba SOC $\lambda = 2\alpha \sqrt{m_F m_S}/ \hbar ^2$. As we present trends for a large
parameter space, unless otherwise specified, we will consider the case for $m_F=m_S=m$ and $k_F=q_F$. 

In Figs.~\ref{fig:TF2}(a) and (b) we show the conductance ratio between the spin-flip and conventional AR, 
$\bar{G}_{\rm{AR}}/G_{\rm{AR}}$, our proxy for singlet and triplet interfacial pairing, as function of the barrier strength and SOC. Remarkably, 
$\bar{G}_{\rm{AR}}/G_{\rm{AR}} >>1$, even for a small spin polarization, $P=0.2$, a nearly complete triplet pairing is possible, $> 90\, \%$ ($96 \,  \%$)
for in-plane (out-of-plane) $\bm{M}$.
A striking enhancement of the triplet contribution is feasible for a wide range of barrier strengths, accompanied with a suitable SOC. As shown
in Fig.~\ref{fig:TF2}, the triangle region of this dominance increases considerably for a larger $P=0.7$ and it is approximately
delimited with lines T1 and T2,
\begin{equation}
\text{T1:}  \:  \: \lambda=2Z/\sqrt{1-P}, \quad \quad \text{T2:} \: \: \lambda=2Z,
\label{eq:L12}
\end{equation}
excluding the half-metals, $P=1$.
Our findings suggest that even simple $s$-wave junctions with only one magnetic region of a small $P$ and interfacial SOC can support 
robust spin-triplet currents. These trends are also preserved for an out-of-plane $\bm{M}$ [Figs.~\ref{fig:TF2}(a), (b) inset]. 

To explore this peculiar behavior and the origin of the triangle region with enhanced triplet pairing, in Fig.~\ref{fig:TF2}(c) we consider
the total $G$ for $P=0.7$ showing  G1 and G2 which denote local maxima in $G$. 
This high-$G$ region, delimited by G1,2, shows a similarity, but not complete overlap with the enhanced triplet region. Such a relatively high-subgap $G$ is in contrast 
to the common expectation that for a strong barrier ($Z>>1$) normal metal/S (N/S) junction would resemble a tunnel contact with a small interfacial 
transparency $T=1/(1+Z^2) \ll 1$~\cite{Blonder1982:PRB}.

For highly-polarized F region, $P=0.7$, conventional AR is strongly
suppressed~\cite{Zutic2004:RMP,Soulen1998:S}. $G$ for such F/S junction should be even lower than for the N/S counterpart with the same large $Z$. 
A striking discrepancy with these expectations comes from the neglect of the SOC and unconventional AR. Even  for a 
strongly-polarized F region, high $G$ is compatible with large $Z$ and  strong SOC. 
In the opposite regime of no SOC ($\lambda \rightarrow 0$), the triplet component will vanish [Fig.~\ref{fig:TF2}(b)], but
there is still a region with only small SOC, $\lambda \sim 0.5$, and a large triplet pairing.

In Fig.~\ref{fig:TF2}(d) we resolve $G$ for four AR processes, responsible for proximity effects,
to examine the evolution of relative contribution of singlet and triplet pairing
with interfacial parameters. While local maxima of $G$  along G1 
arise from singlet contributions $\left|{\uparrow \downarrow}\right\rangle$, $\left|{\downarrow \uparrow}\right\rangle$
and a tiny minority spin-triplet pairing $\left|{\downarrow \downarrow}\right\rangle$, G2 
occurs from majority spin-triplet pairing $\left|{\uparrow \uparrow}\right\rangle$. 
This opens a path to tailor junctions parameters which would selectively remove the singlet contribution and ensure that
transport properties are dominated by (majority) spin-triplet pairing.

The origin the dominant triplet contribution bounded by the T1 and T2 can be traced to the normal-state properties
in the corresponding F/N junction by taking $\Delta=0$. This is further shown in Supplemental Material (See Ref.~\cite{SM19}).
At the interface (barrier region), the dispersion relation is 
$E = \hbar^2 (k_z^2+k_\parallel^2)/2m-\mu+V_0  \pm \alpha k_\parallel/d$.
The energy band is split due to 
SOC [see Fig.~\ref{fig:TF1}(b)] and shifted up by the barrier potential 
(assuming $V_0>0$, but $V_0<0$ gives the same results).
A spinor of an incident electron with $\bm{k}_\parallel$ can be decomposed into barrier
eigenspinors, $\left | \chi_\sigma \right\rangle  =  \left\langle \chi_+  {\left |\right.}\chi_\sigma \right\rangle 
\left | \chi _+ \right\rangle+\left\langle \chi _-  {\left |\right.} \chi _\sigma \right\rangle \left | \chi _- \right\rangle$,   
$\chi_\pm = (1/\sqrt{2})  \left(\pm e^{i\gamma}, 1 \right)^T$, 
with helicity $\pm1$,  where $\gamma=\tan^{-1}(k_x/k_y)$. 
We recognize that these two helicities for outer/inner band have {\em inequivalent} effective barriers~\cite{SM19}
\begin{equation}
Z^+_{\rm{eff}} = 2Z + {\lambda}{k_\parallel }/{k_F}, \quad Z^-_{\rm{eff}} = 2Z- {\lambda}{k_\parallel }/{k_F}.
\label{eq:Zeff}
\end{equation}
Since $Z, \lambda k_\parallel/{k_F} \geq 0$, for positive helicity the barrier is enhanced, $Z^+_{\rm{eff}} \geq Z$. However,  
for negative helicity, at $Z = \lambda k_\parallel/2{k_F}$, $Z^-_{\rm{eff}}$ becomes effectively completely transparent and
can give a dramatically increased $G$.

\begin{figure}[t]
\centering
\includegraphics*[trim=0.6cm 1.3cm 0.2cm 0.2cm,clip,width=8.5cm]{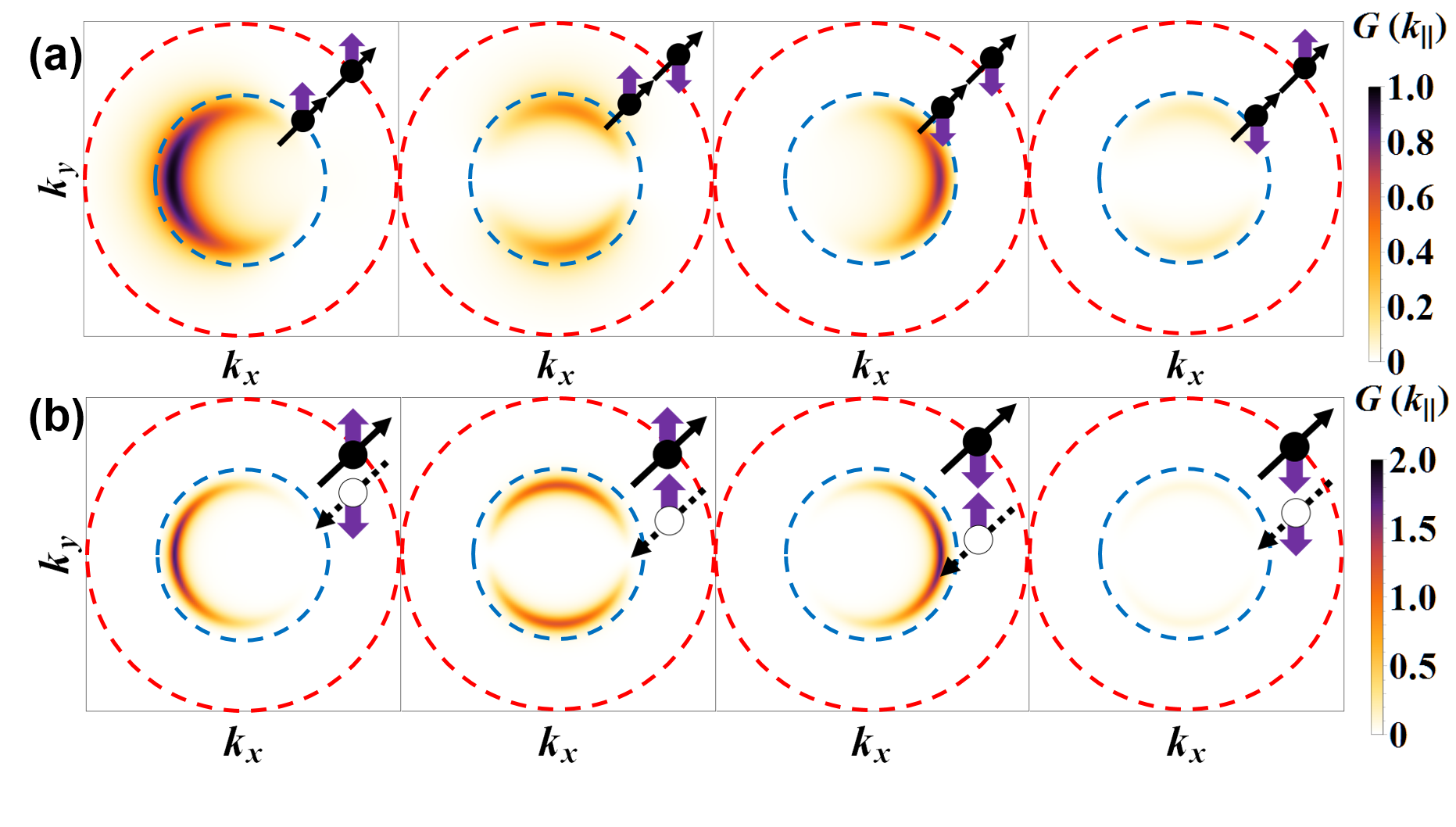}
\caption{
Normalized $\sigma$ and $k_\parallel$-resolved $G$ for different scattering processes in (a) F/N  and  (b) F/S junctions with $P = 0.7$, $Z = 4.4$,  
$\lambda= 20$, $\bm{M} \parallel -y$ . No spin flip in the first and third panels.
The red (blue) circle has a radius $k_F$ ($k_ \downarrow$) of the spin-averaged and spin-down Fermi wave 
vectors, respectively.
}
\label{fig:TF3}
\end{figure}

The effect of this selective barrier transparency and the resulting open channels for a given $\bm{k}_\parallel$ and $\sigma$, can be
clearly seen in Fig.~\ref{fig:TF3}(a). The dominant contribution to $k_\parallel$-resolved conductance comes from the open channels 
located on the circle of radius $k_\parallel= (2Z/\lambda)k_F$. To maximize $G$ for the F/N junction, we can identify several contributing factors. 
(i) The number of open channels, $N(Z,\lambda)$,  should be large. Located on the circle of radius 
$k_\parallel=(2Z/\lambda)k_F$, their number increases with the perimeter, $N(Z,\lambda) \propto k_\parallel$.
(ii) The open channels should exclude evanescent waves for large $k_\parallel$, not contributing to $G$.
This range of $k_\parallel$ follows from the Snell's law~\cite{Zutic2000:PRB}, for incident $\downarrow$ ($\uparrow$) 
electron: $k_\parallel \leq k_\downarrow$ ($k_\parallel \leq k_F$). In the extreme cases, $k_\parallel\equiv(2Z/\lambda)k_F=k_\downarrow$
and $k_\parallel\equiv(2Z/\lambda)k_F=k_F$, we recover exactly T1 and T2 from Eq.~(\ref{eq:L12}). (iii) With spin-momentum locking of interfacial
helical states, an enhanced F/N transmission depends also on the spin matching with the incident spin~\cite{SM19}, in addition to the usual wave vector 
matching~\cite{Zutic1999:PRBa}. 

From these considerations we can understand why, instead of having full circles of open channels, in Fig.~\ref{fig:TF3} we see
crescent-like shapes with completely open channels only for both spin and $k_\parallel$ matching. This picture can be verified
from a simple, but accurate, analytical description of F/N transmission using selective junction transparency~\cite{SM19}.
The transmission decomposed into spin-conserving and spin-flip part, 
$T_\sigma= T_{\sigma \sigma} + T_{\sigma  -\sigma}$, yields
\begin{equation}
T_{\sigma \sigma} \propto \left[1 - \sigma \cos \left(\gamma  + \phi \right) \right]^2, \quad
T_{\sigma -\sigma} \propto \sin^2 \left(\gamma  + \phi \right),
\label{eq:shape}
\end{equation}
confirming  $\pi/2$ and $\pi$ symmetry from Fig.~\ref{fig:TF3}(a), respectively. Here previously given 
angles $\phi$ and $\gamma$ describe the in-plane orientation of $\bm{M}$ and the barrier eigenspinor.

This analysis applies also to F/S junctions, revealing in Fig.~\ref{fig:TF3}(b) a similar angular dependence of $k_\parallel$-resolved $G$
due to conventional and spin-flip AR. Some quantitive modifications from the F/N case, can be understood  already 
without SOC due to a different condition for a perfect F/S transparency at normal incidence were all the wave vectors 
can be unequal $k_\uparrow k_\downarrow=q_F^2$~\cite{Zutic1999:PRBa,Zutic2000:PRB}. For F/S junctions the 
condition for open channels again requires $k_\parallel \leq k_F$ which excludes the evanescent states in AR. The only 
subtlety is $G_{\uparrow \uparrow}$ from spin-flip AR where we could expect that $k_F< k_\parallel \leq k_\uparrow$ is also
possible. However, such a large $k_\parallel$ would result in a strongly decaying wave vector in the S region [recall the 
expression for $q^{e\,(h)}$] with its inverse smaller than the BCS
coherence length and thus render ineffective 
any contribution for spin-majority pairing with $k_\parallel \leq k_F$. This provides a guidance for a choice of
junction parameters giving an enhanced spin-triplet paring between the lines T1 and T2 in Eq.~(5), even for previously
unexpected regimes with only a small $P$.

\begin{figure}[t]
\centering
\includegraphics*[trim=0.3cm 0.35cm 0.0cm 0.3cm,clip,width=8.5cm]{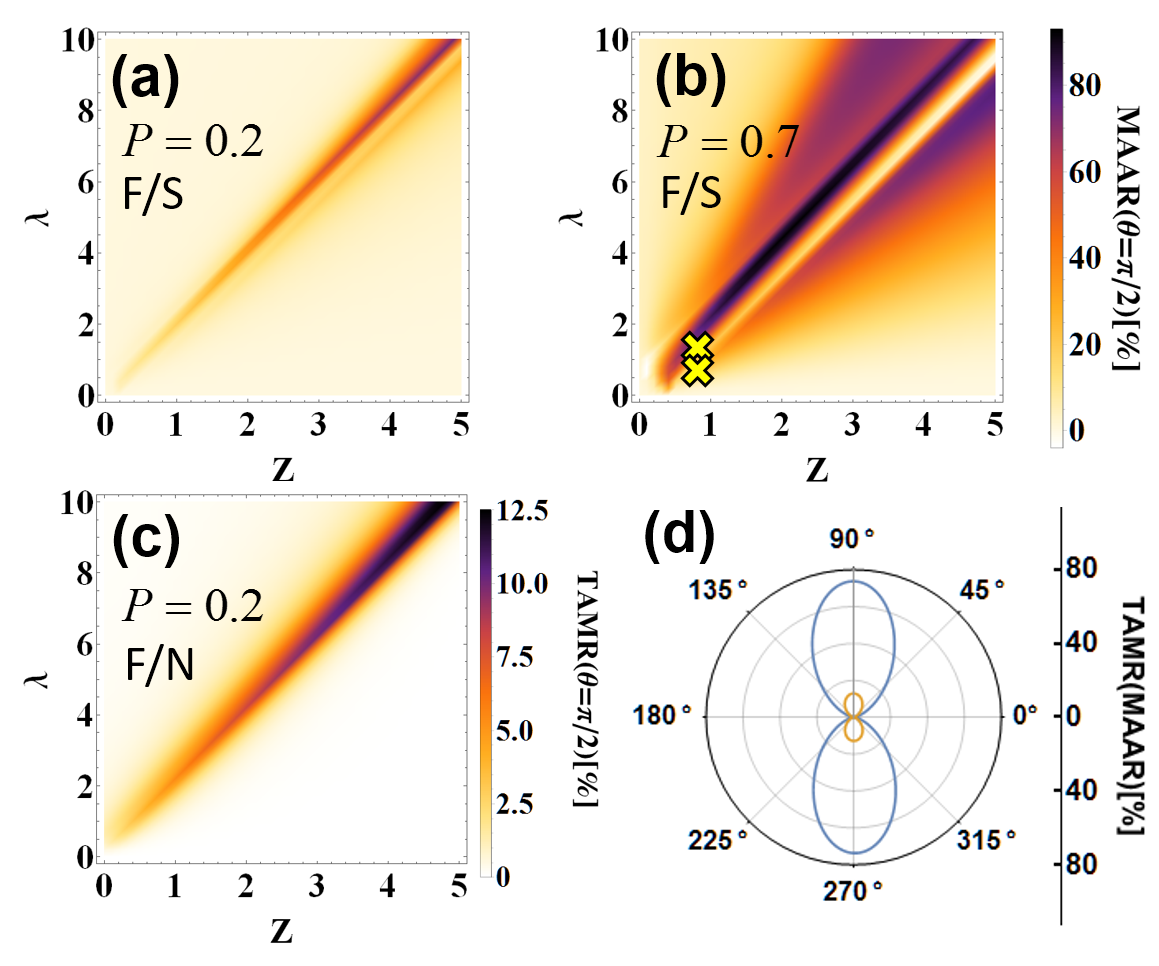}
\caption{
Amplitude of out-of-plane magnetoanistotropic Andreev reflection (MAAR) as a function of  interface parameters $Z$ and $\lambda$ for (a) $P = 0.2$ and (b) 
$P = 0.7$. (c) The corresponding tunneling anisotropic magnetoresistance  (TAMR) when superconducting gap vanishes for $P = 0.2$. (d) A comparison 
between out-of-plane TAMR (yellow) and MAAR (blue), $P = 0.2$, $Z = 5$ and $\lambda=10.2$.
}
\label{fig:TF4}
\end{figure}

In addition to directly measuring the spin structure of $G$ 
or spin current, an experimental test of our
predictions for enhanced spin-triplet pairing could be realized through probing magnetic anisotropy of conductance 
in F/S junctions, referred to as magnetic anisotropic Andreev reflection (MAAR)~\cite{Hogl2015:PRL}. 
MAAR and it is better studied normal-state analog, tunneling anisotropic magnetoresistance (TAMR)~\cite{Moser2007:PRL,Fabian2007:APS}, 
can be expressed for out-of-plane rotation of $\bm{M}$  [Fig.~\ref{fig:TF1}(a)] as~\cite{Hogl2015:PRL}
\begin{equation}
\mathrm{TAMR(\theta), \, MAAR(\theta)}=[G(0)-G(\theta)]/G(\theta),
\label{eq:MAAR}
\end{equation}
where angle $\theta$ is between $\bm{M}$ and the interface normal. 
From the evolution of MAAR, 
shown in Figs.~\ref{fig:TF4}(a) and (b) for $P=0.2$ and $P=0.7$, we see that it closely follows the trends of the enhanced majority spin-triplet pairing from 
Figs.~\ref{fig:TF2}(a) and (b). It is this spin-triplet component that is responsible for a large increase of MAAR compared to TAMR,
in the normal state, Figs.~\ref{fig:TF4}(a),\, (c), \,  (d). Even for $P=0.2$ the resulting increase can reach an order of magnitude and become
much larger for $P=0.7$ where it was recently measured in all-epitaxial Fe/MgO/V junctions~\cite{Martinez2018:P} to exceed 1000! 
Rather than change MAAR to TAMR by increasing the temperature above the 
critical temperature (for vanadium $\sim 4$ K), experimentally it is more convenient to 
reach the normal state by increasing the  bias,  $V > \Delta$ at a fixed temperature~\cite{Martinez2018:P}. 

Such Fe/MgO/V junctions simplify the analysis of the observed magnetic anisotropy since they have two stable 
zero-field ($B=0$) states with mutually orthogonal $\bm{M}$: in-plane and out-of-plane~\cite{Martinez2018:P,Martinez2018:SR}. 
This removes common complications in other F/S junction by decoupling the influence of the $B$-field required for 
rotating $\bm{M}$ which could alter the magnitude of magnetic anisotropy and create spurious effects from vortices. 
Junction parameters $Z=0.83$ ($V_0=0.3$ eV, d=17 nm), $\lambda= 0.79$, 1.44 ($\alpha=5.5$ eV\AA$^2$), 
describing two measured Fe/MgO/V samples with MAAR of 10-20 \% 
(TAMR  only $\sim$0.01$\,$\%)~\cite{Martinez2018:P} are marked  in Fig.~\ref{fig:TF4}(b). This small SOC, $\lambda \sim1$, smaller
than in Fe/GaAs/Au TAMR studies~\cite{Moser2007:PRL}, is already sufficient for a dominant triplet pairing. 

While we employ a simple approach which naturally suggests a number of generalizations, from inclusion of the self-consistent pair potential, finite $B$-fields, 
study of critical temperature, or more complex barrier description~\cite{Wu2014:PRB,Miyoshi2005:PRB,Simensen2018:PRB,Barsic2009:PRB,
Valls2010:PRB,Costa2017:PRB}, 
its transparency already reveals several important trends and can support peculiar experimental observation of a giant MAAR~\cite{Martinez2018:P}. 
Our implications for enhanced triplet pairing and MAAR detection could also be relevant for 
two-dimensional materials, as supported by the work in Refs.~\cite{Beiranvand2016:PRB,Lv2018:PRB}.
Another extension of this work could include the role of magnetic textures which themselves result in synthetic spin-orbit coupling and could be used to control
Majorana bound states~\cite{Desjardins2019:P,Fatin2016:PRL,Matos-Abiague2017:SSC,Klinovaja2013:PRL,Yang2016:PRB,Zhou2019:PRB,
Gungordu2018:PRB,Mohanta2019:P}.

Similar to the advances in realizing large magnetoresistive effect, not by employing complex ferromagnets with nearly complete spin polarization, but rather choosing a
suitable nonmagnetic barrier~\cite{Parkin2004:NM,Yuasa2004:NM}, our findings suggest what could constitute a 
suitable interface to realize enhanced spin-triplet proximity. In particular, to further enhance such triplet pairing with only a very small spin polarization 
of a ferromagnet, a challenge would be to design interfaces which could simultaneously provide a large
spin-orbit coupling and large potential barrier. 

We thank Petra H{\"o}gel and Farkhad Aliev for valuable discussions. This work is supported by 
Department of Energy, Basic Energy Sciences Grant DE-SC0004890 and the UB Center
for Computational Research. 

\bibliography{Triplet}
\end{document}